\documentclass[aps,pra,twocolumn,showpacs]{revtex4}

\usepackage[dvips]{graphicx}

\input{epsf}

\begin{document}

\title{Interacting fermions in quasi-one-dimensional harmonic traps}

\author{G. E. Astrakharchik$^{(1,2)}$, D. Blume$^{(3)}$, S. Giorgini$^{(1)}$, and L. P. Pitaevskii$^{(1)}$}
\address{$^{(1)}$Dipartimento di Fisica, Universit\`a di Trento and BEC-INFM, I-38050 Povo, Italy\\
$^{(2)}$Institute of Spectroscopy, 142190 Troitsk, Moscow region, Russia\\
$^{(3)}$Department of Physics, Washington State University,
Pullman, Washington 99164-2814, USA}

\date{\today}

\begin{abstract}
Quasi-one-dimensional (quasi-1d) two-component Fermi gases with effectively attractive and repulsive interactions
are characterized for arbitrary interaction strength. The ground-state properties of the gas confined in highly 
elongated harmonic traps are determined within the local density approximation. For strong attractive effective 
interactions the existence of a molecular Tonks-Girardeau gas is predicted. The frequency of the lowest breathing 
mode is calculated as a function of the coupling strength for both attractive and repulsive interactions.  
\end{abstract}


\maketitle

The study of cold quasi-1d atomic quantum gases presents 
a very active area of research. So far, most of the 
experimental \cite{EXP1} and 
theoretical \cite{Olshanii,Petrov,Dunjko,Menotti,Girardeau2} investigations
have been devoted to quasi-1d Bose
gases and, in particular, to the strongly-interacting 
Tonks-Girardeau gas, which can be mapped 
to a gas of non-interacting 
fermions~\cite{Girardeau1,Olshanii,Reichel}. 
Quasi-1d atomic Fermi gases have not been realized experimentally yet.

The role of interactions in quasi-1d atomic Fermi gases has been studied mainly in 
connection with Luttinger liquid theory \cite{Wonneberger,Recati}. 
Recati {\it et al.}~\cite{Recati} 
investigate the properties of a two-component Fermi gas with {\em{repulsive}} 
interspecies interactions confined in highly-elongated harmonic traps. In the limit of
weak and strong coupling these authors relate the parameters of the Luttinger Hamiltonian, 
which describe the low-energy properties of the gas, to the microscopic parameters of the 
system. The Luttinger model is used to analyze the manifestations of the uncoupled dynamics
of spin and density waves (spin-charge separation).  

This Letter investigates the properties of inhomogeneous quasi-1d 
two-component Fermi 
gases under harmonic confinement with 
{\em{attractive and repulsive}} interspecies interactions.
The present study is based on the exact equation of state of a homogeneous 1d system of 
fermions with zero-range attractive \cite{Gaudin,Krivnov} and repulsive \cite{Yang} 
interactions treated within the local density approximation (LDA). We calculate the energy per particle, the size 
of the cloud, and the frequency of the lowest compressional mode as a function of the 
effective 1d coupling constant, including infinitely strong attractive and repulsive interactions. 
Moreover, for attractive interactions we 
discuss the cross-over from the weak- to the strong-coupling 
regime and point out the possibility of forming a mechanically stable molecular Tonks-Girardeau gas.

Quasi-1d two-component Fermi gases with effectively attractive and repulsive 1d interspecies interactions
can be realized in highly-elongated traps. The behavior of atomic gases tightly-confined in two 
directions can, if the confinement is chosen properly, be characterized to a very good approximation by 
an effective 1d coupling constant, $g_{1d}$, which encapsulates the atom-atom interaction strength. This 
coupling constant can be tuned to essentially any value, including zero and $\pm\infty$, by varying the 
3d $s$-wave scattering length $a_{3d}$ through application of an external magnetic field in the proximity 
of a Feshbach resonance. 
 
In homogeneous 1d Fermi gases with attractive interactions, 
sound waves propagate with 
a well defined velocity, while
spin waves exhibit a gap \cite{Krivnov}. Furthermore, in the strong-coupling 
regime, the ground state is comprised of bosonic molecules, consisting of two fermions with different 
spin whose spatial size is much smaller than the average intermolecular distance \cite{Krivnov}. 
Consequently, BCS-type equations have been discussed for effectively attractive 1d interactions 
\cite{1DBCS}. The quasi-1d molecular Bose gas discussed here has similarities with the formation of a 
molecular Bose-Einstein condensate (BEC) from a 3d Fermi sea close to a magnetic atom-atom Feshbach 
resonance \cite{EXP3}, which may allow study of the BCS-BEC cross-over \cite{BCSBEC}. In contrast to 
the 3d case, the cross-over from the weak-coupling regime of large, overlapping Cooper pairs to the 
strong-coupling regime of small, tightly bound Bose pairs can be investigated in 1d within an exactly 
solvable model. 

Consider a two-component atomic Fermi gas confined in a  
highly-elongated trap. The fermionic atoms are assumed to
belong to the same atomic species, that is, to have the same mass $m$, but
to be
trapped in different hyperfine states $\sigma$, 
where $\sigma$ represents a generalized spin 
or angular momentum, $\sigma=\uparrow$ or $\downarrow$. The trapping potential is assumed to be
harmonic and axially symmetric,
\begin{equation}
V_{trap}=\sum_{i=1}^N\frac{1}{2}m\left( \omega_\rho^2\rho_i^2 + 
\omega_z^2 z_i^2 \right) \;.
\label{trap}
\end{equation} 
Here, $\rho_i=\sqrt{x_i^2+y_i^2}$ and $z_i$ denote, respectively, 
the radial and longitudinal 
coordinate of the $i$th atom; $\omega_{\rho}$ and $\omega_z$ 
denote, respectively, the angular frequency in the 
radial and longitudinal direction; and $N$ denotes the total number of atoms. 
We require that the anisotropy parameter $\lambda \ll 1$,
where $\omega_z=\lambda \omega_\rho$, so that 
the transverse motion is ``frozen'' to zero point oscillations. 
At zero temperature this implies that
the Fermi energy associated with the 
longitudinal motion of the atoms in the absence of interactions, $\epsilon_F=N\hbar\omega_z/2$, is much 
smaller than the separation between the
levels in the transverse direction, $\epsilon_F\ll\hbar\omega_\rho$. 
This
condition is fulfilled if $N\lambda\ll 1$. 
The outlined scenario can be realized experimentally with
present-day technology using optical traps.

\begin{figure}
\begin{center}
\includegraphics*[width=7cm]{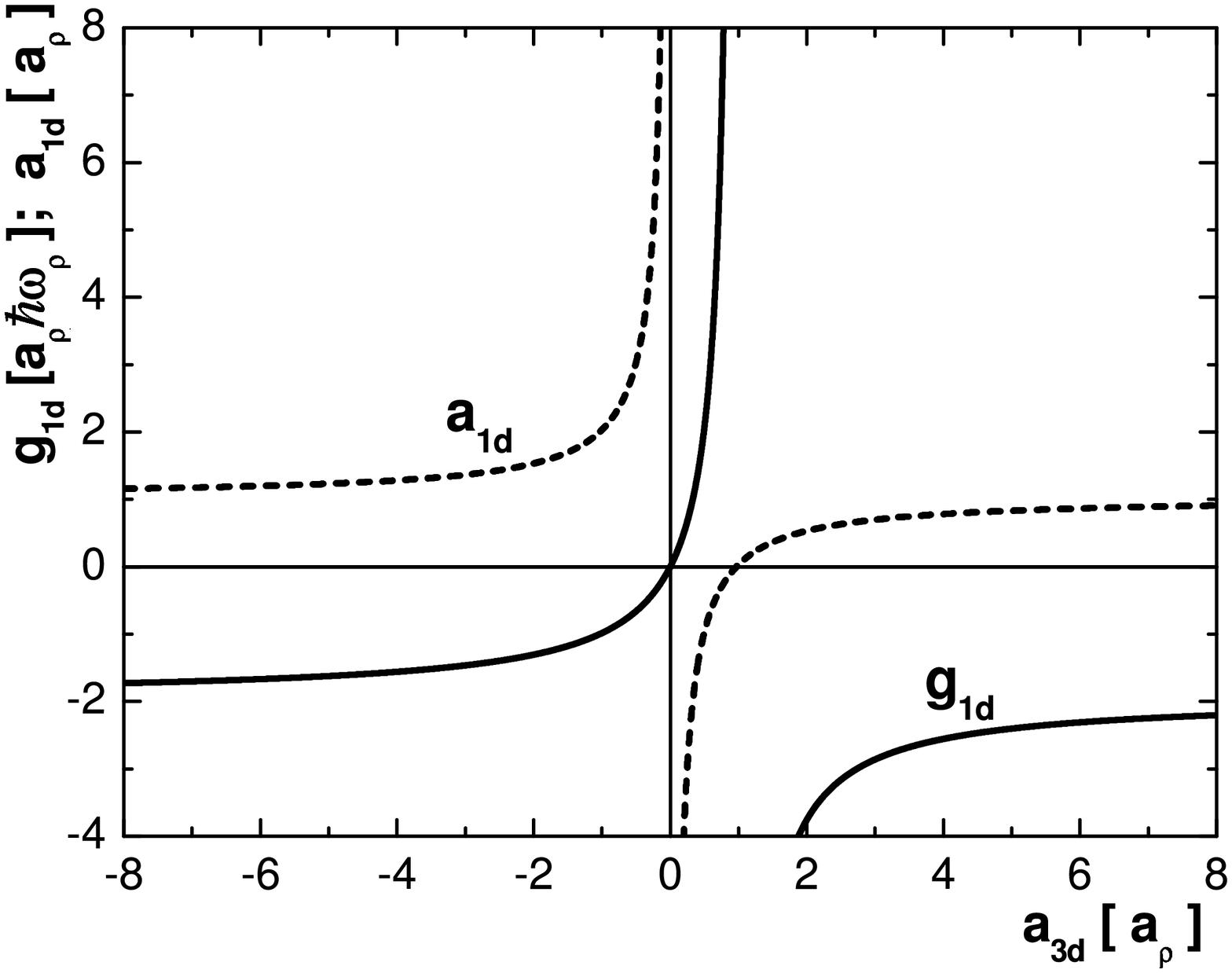}
\caption{Effective 1d coupling constant $g_{1d}$ 
[solid line, Eq.~(\protect\ref{g1d})], together with 
effective 1d scattering length
$a_{1d}$ [dashed line, Eq.~(\protect\ref{a1d})] as a function of $a_{3d}$.}
\label{fig1}
\end{center}
\end{figure}

If the Fermi gas is kinematically in 1d, it can be described
by an effective 1d Hamiltonian with contact interactions,
\begin{equation}
H=N\hbar\omega_\rho + H_{1d}^0 + \sum_{i=1}^N\frac{1}{2}m\omega_z^2z_i^2 \;,
\label{hamiltonian1}
\end{equation}   
where
\begin{equation}
H_{1d}^0=-\frac{\hbar^2}{2m}\sum_{i=1}^N\frac{\partial^2}{\partial z_i^2} + g_{1d}\sum_{i=1}^{N_\uparrow}
\sum_{j=1}^{N_\downarrow}\delta(z_i-z_j) 
\label{hamiltonian2}
\end{equation}
and $N=N_{\uparrow}+N_{\downarrow}$.
This effective Hamiltonian 
accounts for
the interspecies atom-atom interactions, 
which are parameterized by the 3d $s$-wave scattering length $a_{3d}$,
through 
the effective 1d coupling constant $g_{1d}$~\cite{Olshanii}, 
\begin{equation}
g_{1d}=\frac{2\hbar^2a_{3d}}{m a_\rho^2} \frac{1}{1-A a_{3d}/a_\rho} \;,
\label{g1d}
\end{equation}
but neglects the typically much weaker
intraspecies $p$-wave interactions.
In Eq.~(\ref{g1d}), 
$a_\rho=\sqrt{\hbar/m\omega_\rho}$ is the characteristic oscillator length 
in the transverse direction and 
$A=|\zeta(1/2)|/\sqrt{2}\simeq 1.0326$. Alternatively, $g_{1d}$ can be expressed through the effective 1d scattering length
$a_{1d}$, $g_{1d}=-2\hbar^2/(ma_{1d})$, where
\begin{equation}
a_{1d}=-a_\rho\left(\frac{a_\rho}{a_{3d}}-A\right) \;.
\label{a1d}
\end{equation}
Figure~\ref{fig1} shows $g_{1d}$ and $a_{1d}$ as a 
function of the 3d $s$-wave scattering length $a_{3d}$,
which can be varied continuously by application of an external field.
The effective 1d interaction is repulsive, $g_{1d}>0$, 
for $0<a_{3d}<a_{3d}^c$ ($a_{3d}^c=0.9684 a_\rho$), and  
attractive, $g_{1d}<0$, for
$a_{3d}>a_{3d}^c$ and for $a_{3d}<0$. 
By varying $a_{3d}$, it is possible to 
go adiabatically from the strongly-interacting repulsive regime 
($g_{1d}\to +\infty$ or $a_{3d}\lesssim a_{3d}^c$), through the weakly-interacting
regime ($g_{1d}\sim 0$) to the strongly-interacting
attractive regime ($g_{1d}\to -\infty$ or $a_{3d}\gtrsim a_{3d}^c$).

For two fermions with different spin
the Hamiltonian 
$H_{1d}^0$, Eq.~(\ref{hamiltonian2}), 
supports one bound state
with binding energy 
$\epsilon_{bound}=-\hbar^2/(ma_{1d}^2)$ and spatial extent $\sim a_{1d}$
for $g_{1d}<0$, 
and no bound state for $g_{1d}>0$, that is,
the molecular state 
becomes exceedingly weakly-bound and spatially-delocalized
as $g_{1d}\to 0^-$. In the following we investigate the properties 
of a gas with $N$ fermions, $N_{\uparrow}= N_{\downarrow}$, for both
effectively {\em{attractive and repulsive}} 1d interactions {\em{with
and without}}
longitudinal confinement.

To start with, consider the Hamiltonian $H_{1d}^0$, Eq.~(\ref{hamiltonian2}), 
which describes a homogeneous 1d two-component Fermi gas.
The ground state energy $E_{hom}$ of $H_{1d}^0$ has been
calculated exactly using Bethe's ansatz for both 
attractive \cite{Gaudin} and repulsive \cite{Yang} interactions,
and can be expressed in terms of the linear number density $n_{1d}$,
$n_{1d}=N/L$, where $L$ is the size of the system,
\begin{equation}
\frac{E_{hom}}{N}=\frac{\hbar^2n_{1d}^2}{2m}e(\gamma) \;.
\label{homenergy1}
\end{equation}
The dimensionless parameter $\gamma$ is
proportional to the coupling constant $g_{1d}$, 
$\gamma=m g_{1d}/(\hbar^2n_{1d})$, while
its absolute value
is inversely proportional to the
1d gas parameter $n_{1d}|a_{1d}|$, $|\gamma|=2/n_{1d}|a_{1d}|$. 
The 
function $e(\gamma)$ is obtained by solving a set of 
integral equations \cite{web}, which is
similar to that derived by Lieb and Liniger 
\cite{LL} for 1d bosons with repulsive contact interactions.
To obtain the energy per particle, Eq.~(\ref{homenergy1}), 
we solve
%
these integral equations for attractive $(\gamma<0)$~\cite{Gaudin}
and repulsive interactions ($\gamma>0$)~\cite{Yang}.

\begin{figure}
\begin{center}
\includegraphics*[width=7cm]{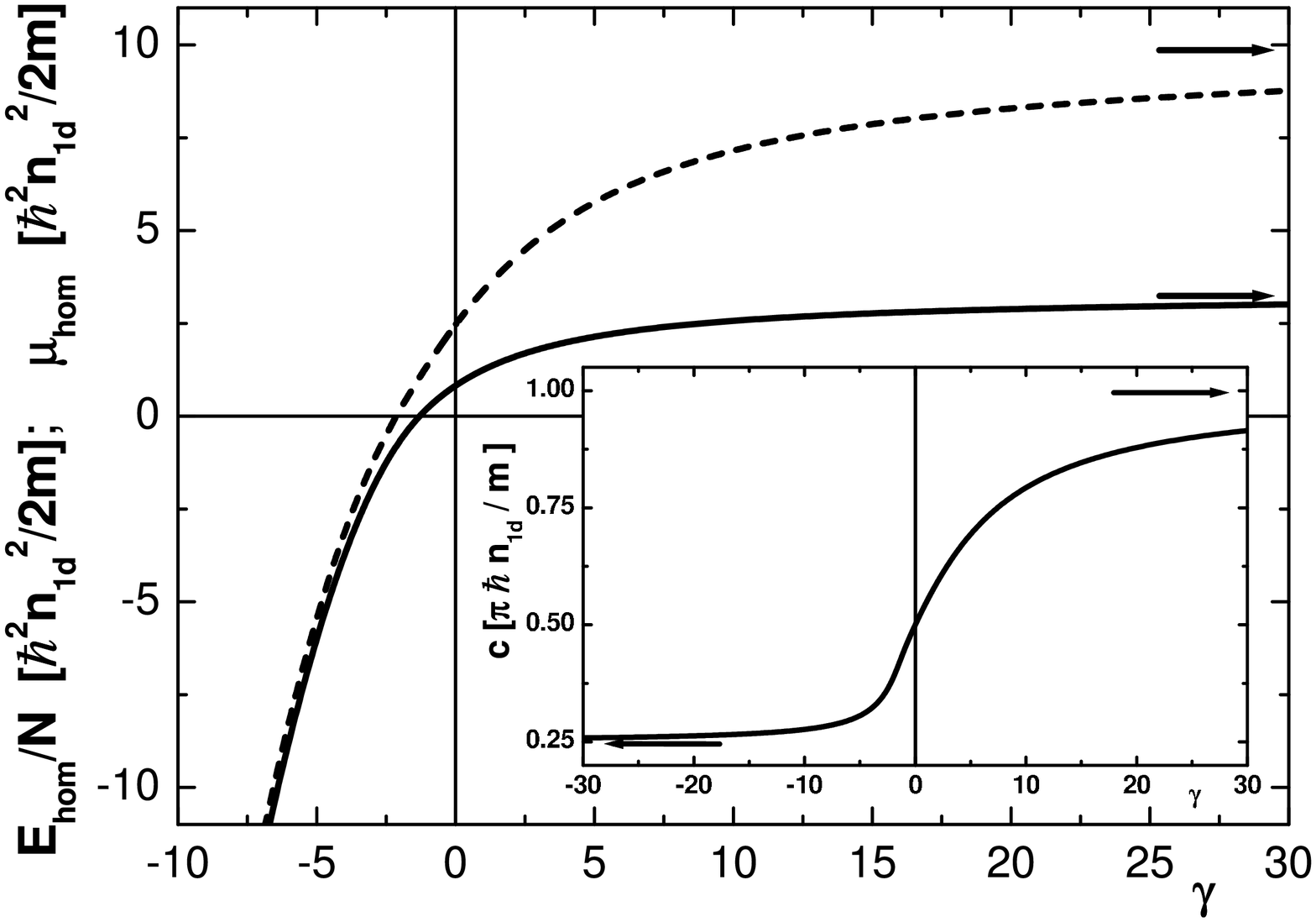}
\caption{$E_{hom}/N$ (solid line), $\mu_{hom}$ (dashed line)
and $c$ (inset) for a homogeneous two-component 1d Fermi gas
as a function of $\gamma$ (horizontal arrows indicate the asymptotic
values of $E_{hom}/N$, $\mu_{hom}$ and $c$, respectively).}
\label{fig2}
\end{center}
\end{figure}

Figure~\ref{fig2} shows 
the energy per particle, $E_{hom}/N$ (solid line), the chemical potential
$\mu_{hom}$, 
$\mu_{hom}=dE_{hom}/dN$ 
(dashed line), and the velocity of sound $c$ (inset), which
is obtained from the compressibility of the system, 
$mc^2=n_{1d}\partial\mu_{hom}/\partial n_{1d}$, as a 
function of the interaction strength $\gamma$. In the weak coupling limit, 
$|\gamma|\ll 1$, 
$\mu_{hom}$ is given by
\begin{equation}
\mu_{hom}=\frac{\pi^2}{4} \; \frac{\hbar^2n_{1d}^2}{2m}+
\gamma \; \frac{\hbar^2 n_{1d}^2}{2m} + 
\cdots\;,
\label{limit1}
\end{equation} 
where the first term on the right hand side is the energy of
an ideal two-component atomic Fermi gas, and the se\-cond term is the
mean-field energy, which 
accounts for interactions. 
The chemical potential increases with increasing $\gamma$,
and reaches an asymptotic value
for $\gamma \rightarrow \infty$~\cite{Recati} (indicated by a horizontal 
arrow in Fig.~\ref{fig2}),
\begin{equation}
\mu_{hom}= \pi^2 \; \frac{\hbar^2n_{1d}^2}{2m}-
\frac{16 \pi^2 \ln(2)}{3 \gamma} \; \frac{\hbar^2n_{1d}^2}{2m} + \cdots\;.
\label{limit2}
\end{equation}
The first term on the right hand side coincides with
the chemical potential of a one-component ideal 1d Fermi gas with 
$N$ atoms. Interestingly,
for $\gamma\gg 1$, the strong atom-atom repulsion between
atoms with different 
spin plays the role 
of an effective Pauli principle.

For attractive interactions and large enough $|\gamma|$
the energy per particle
is negative  (see Fig.~\ref{fig2}), 
reflecting the existence of a molecular Bose gas, which
consists of
$N/2$ 
diatomic molecules with binding 
energy $\epsilon_{bound}$. Each molecule
is comprised of two atoms with different spin.
In the limit $\gamma \rightarrow -\infty$,
the chemical potential becomes
\begin{equation}
\mu_{hom}=-\frac{\hbar^2}{2ma_{1d}^2}+
\frac{\pi^2}{16} \; \frac{\hbar^2n_{1d}^2}{2m}+
\frac{\pi^2}{12 \gamma} \;
\frac{\hbar^2n_{1d}^2}{2m} + \cdots \;.
\label{limit3}
\end{equation}    
The first term is simply $\epsilon_{bound}/2$, one half of 
the binding energy of the 1d molecule, while the second term
is identical to
the chemical potential of a bosonic Tonks-Girardeau gas with
density $n_{1d}/2$, consisting
of $N/2$ molecules with mass 
$2m$ \cite{note}. Importantly,
the compressibility remains positive
for $\gamma \rightarrow -\infty$
[a horizontal arrow in the inset of Fig.~\ref{fig2}
indicates the asymptotic value
of $c$, $c=\pi\hbar n_{1d}/(4m)$], which implies that
two-component 1d Fermi gases 
are mechanically stable even in the strongly-attractive regime.
In contrast, 
the ground state of 1d Bose gases with $g_{1d}<0$
has negative 
compressibility~\cite{McGuire} and is hence mechanically unstable.
The formation of a quasi-1d molecular gas discussed here has some analogies
to the formation
of a 3d molecular BEC~\cite{EXP4}
and
the BCS-BEC 
cross-over for 3d Fermi systems~\cite{BCSBEC}.

Using the solutions for the homogeneous two-component 1d Fermi gas, 
\begin{figure}
\begin{center}
\includegraphics*[width=7cm]{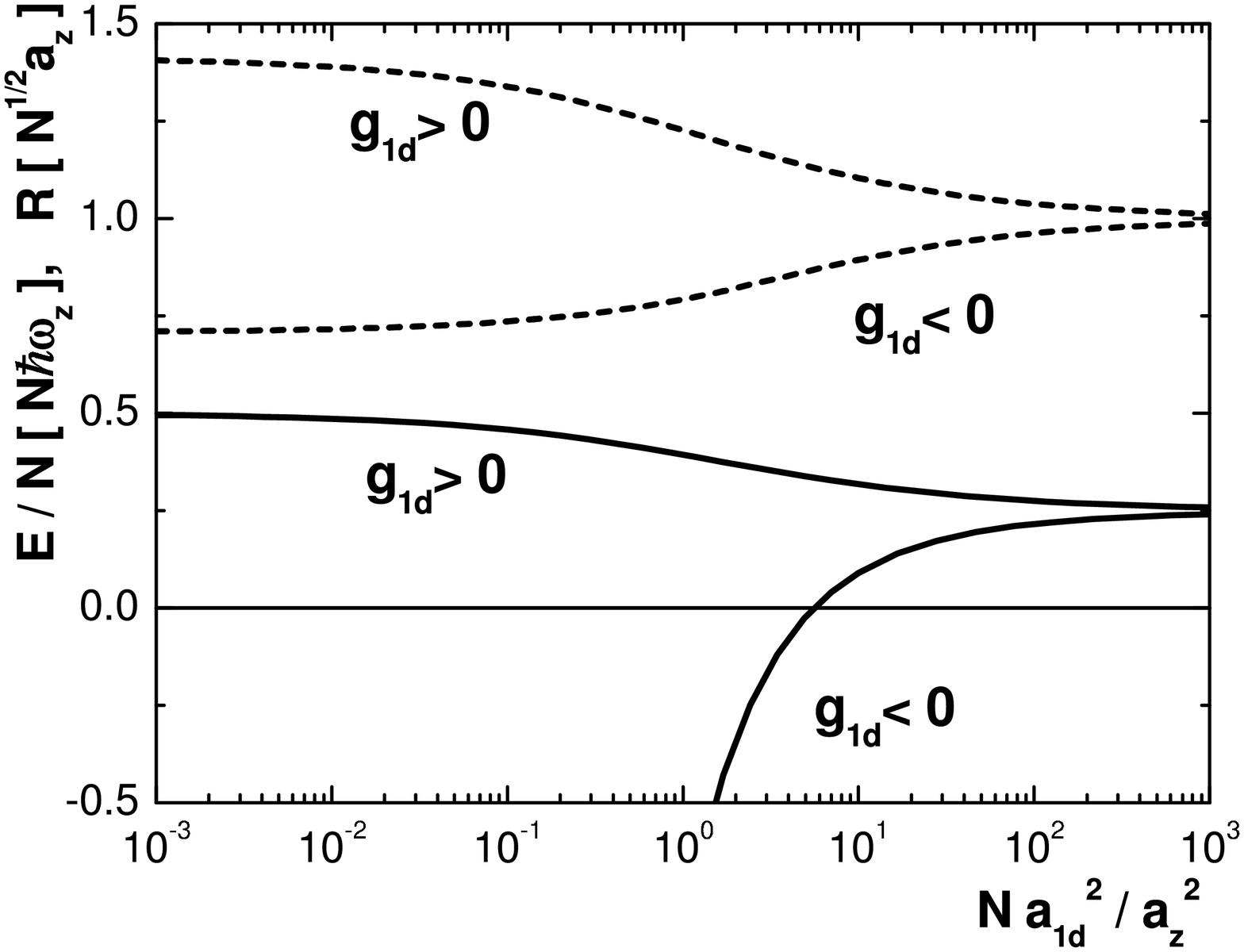}
\caption{Energy per particle, $E/N$ (solid lines), 
and size of the cloud, $R$ (dashed lines), for an inhomogeneous 
two component 1d Fermi gas as a function of the coupling strength $Na_{1d}^2/a_z^2$ for repulsive ($g_{1d}>0$)
and attractive ($g_{1d}<0$) interactions.}
\label{fig3}
\end{center}
\end{figure}
we now describe the 
inhomogeneous gas, 
Eq.~(\ref{hamiltonian1}), within the LDA~\cite{Dunjko,Menotti,Recati}.
This approximation
is applicable
if the size $R$ of the cloud
is much larger than the harmonic 
oscillator length $a_z$ in the longitudinal direction,
$a_z=\sqrt{\hbar/m\omega_z}$, implying
$\epsilon_F\gg
\hbar\omega_z$ and
$N\gg 1$. The chemical potential $\mu$ of the inhomogeneous
system can be determined from the local equilibrium condition,
\begin{equation}
\mu=\mu_{hom}[n_{1d}(z)]+\frac{1}{2}m\omega_z^2z^2 \;,
\label{lda}
\end{equation}
and the normalization condition $N=\int_{-R}^R  n_{1d}(z)dz$,
where $z$ is measured from the
center of the trap and $R=\sqrt{2\mu^{\prime}/(m\omega_z^2)}$, where $\mu^{\prime}=\mu$ for $g_{1d}>0$ and
$\mu^{\prime}=\mu+|\epsilon_{bound}|/2$ for $g_{1d}<0$. 
The normalization condition can be reexpressed
in terms of the dimensionless chemical 
potential $\tilde{\mu}$ and the dimensionless density $\tilde{n}_{1d}$
[$\tilde{\mu}=\mu^{\prime}/(\hbar^2/2ma_{1d}^2)$ and 
$\tilde{n}_{1d}=|a_{1d}|n_{1d}$], 
\begin{equation}
N\frac{a_{1d}^2}{a_z^2}=
\int_0^{\tilde{\mu}} \frac{\tilde{n}_{1d}(\tilde{\mu}-x)}{\sqrt{x}} dx\;.
\label{normalization2}
\end{equation}
This expression emphasizes that the coupling strength
is determined by 
$Na_{1d}^2/a_z^2$;
$Na_{1d}^2/a_z^2\gg 1$ corresponds to the weak 
coupling and $Na_{1d}^2/a_z^2\ll 1$ to the strong coupling regime,
irrespective of whether the interactions are
attractive or repulsive~\cite{Menotti}.

Figure~\ref{fig3} shows the 
energy per particle $E/N$ and the size $R$ 
of the cloud as a function of the coupling strength 
$Na_{1d}^2/a_z^2$ for positive 
and negative $g_{1d}$ calculated within the LDA for an inhomogeneous 
two-component 1d Fermi gas.
Compared to the non-interacting gas, 
for which $R=\sqrt{N}a_z$,
$R$ increases for repulsive interactions
and decreases for attractive interactions. For $Na_{1d}^2/a_z^2 \ll 1$,
$R$ reaches the
asymptotic value $\sqrt{2N}a_z$ for the strongly repulsive regime,
$g_{1d} \rightarrow +\infty$,
and the value $\sqrt{N/2}a_z$ for the strongly attractive regime, 
$g_{1d} \rightarrow -\infty$.
The shrinking of the cloud for attractive 
interactions reflects the formation of tightly bound molecules.
In the limit $g_{1d} \rightarrow -\infty$, the energy per particle 
approaches $\epsilon_{bound}/2+N \hbar \omega_z/4$, indicating the formation of a
molecular bosonic Tonks-Girardeau gas, consisting of
$N/2$ molecules.

Using a sum rule approach, the frequency $\omega$
of the lowest compressional (breathing) mode of harmonically trapped 1d 
gases can be calculated from the
mean-square size of the cloud $\langle z^2 \rangle$~\cite{Menotti},
\begin{equation}
\omega^2=-2\frac{\langle z^2\rangle }{d\langle z^2\rangle/d\omega_z^2} \;.
\label{collmode}
\end{equation} 
In the weak and strong coupling regime ($Na_{1d}^2/a_z^2 \gg 1$ and $\ll 1$,
respectively), $\langle z^2 \rangle$ has the same 
dependence on $\omega_z$ as the
ideal 1d Fermi gas. Consequently, $\omega$ is in these limits given by
$2\omega_z$, irrespective of whether the interaction is repulsive
or attractive.
%
Solid lines in Fig.~\ref{fig3} show $\omega^2$, determined 
numerically from Eq.~(\ref{collmode}),
as a function of the interaction strength $N a_{1d}^2/a_z^2$.
A non-trivial behavior of $\omega^2$ as a function of
$Na_{1d}^2/a_z^2$ is visible. 
To gain further insight,
we calculate the first correction $\delta\omega$ to the breathing mode 
frequency $\omega$
\begin{figure}
\begin{center}
\includegraphics*[width=7cm]{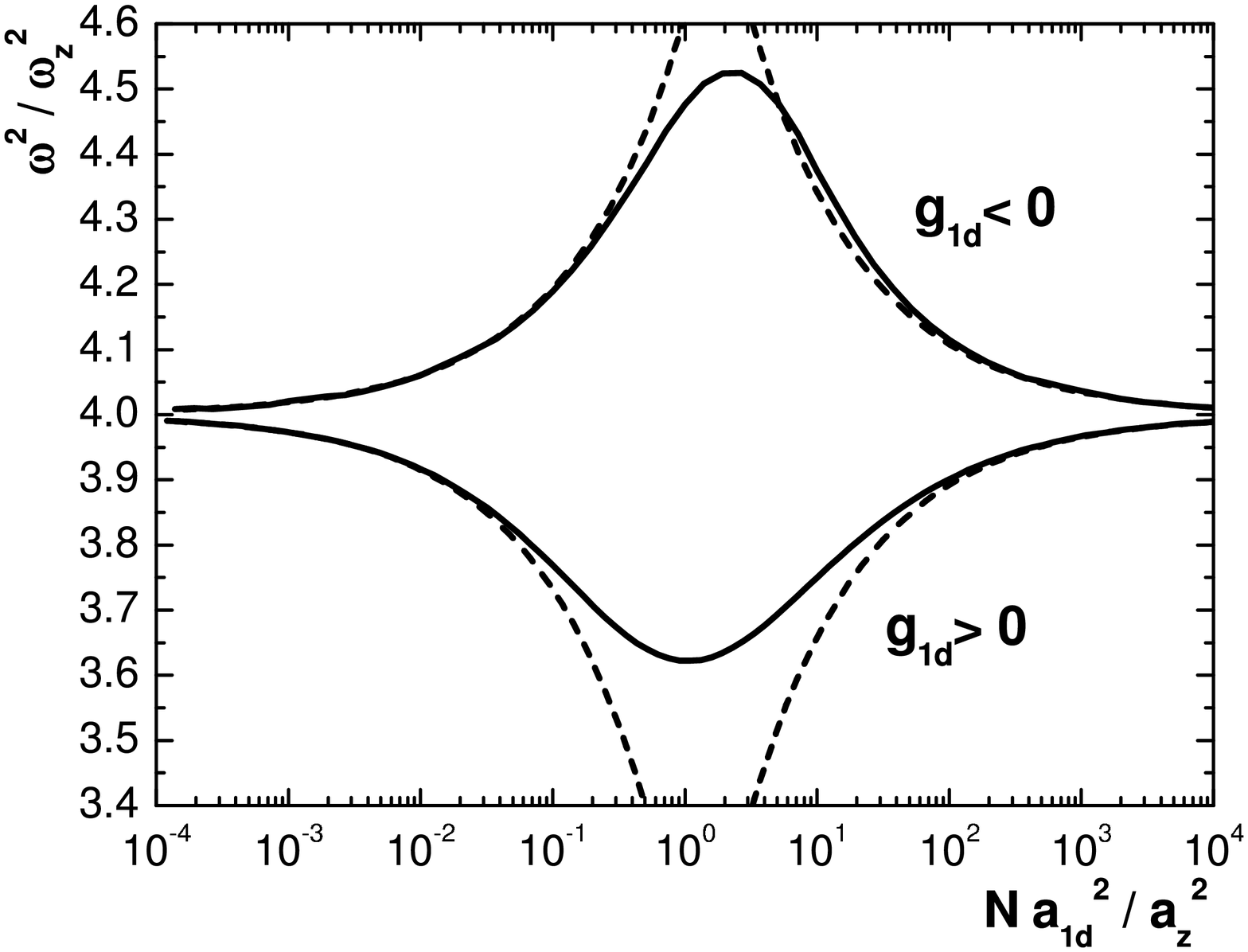}
\caption{Square of the lowest breathing mode frequency,
$\omega^2$,
as a function of the coupling strength $Na_{1d}^2/a_z^2$ for 
an inhomogeneous 
two-component 1d Fermi gas with repulsive 
($g_{1d}>0$) and attractive ($g_{1d}<0$) interactions determined numerically
from Eq.~(\ref{collmode}) (solid lines).
Dashed lines show analytic expansions (see text), which are valid 
in the weak and strong coupling regime.}
\label{fig4}
\end{center}
\end{figure}
[$\omega=2\omega_z(1+\delta\omega/\omega_z+\cdots)$]
analytically for weak repulsive and attractive interactions, as well 
as for strong 
repulsive and attractive interactions.
For the weak coupling regime, we find
$\delta\omega/\omega_z=\pm (4/3\pi^2)/(Na_{1d}^2/a_z^2)^{1/2}$, 
where the minus sign applies to repulsive interactions and
the plus sign to attractive interactions.
For the strong coupling regime, we find
$\delta\omega/\omega_z=-[16\sqrt{2}\ln(2)/15\pi^2](Na_{1d}^2/a_z^2)^{1/2}$
for repulsive interactions and 
$\delta\omega/\omega_z=(8\sqrt{2}/15\pi^2)(Na_{1d}^2/a_z^2)^{1/2}$
for attractive interactions.
Dashed lines in Fig.~\ref{fig4} show 
the resulting analytic expansions for $\omega^2$.
These expansions describe the lowest breathing mode frequency quite
well over a fairly large range of interaction strengths.
Not surprisingly, their validity breaks down for
$Na_{1d}^2/a_z^2\sim 1$.  

In conclusion, we have 
investigated the cross-over from 
weak to strong coupling of quasi-1d harmonically trapped two-component 
Fermi gases with both repulsive and attractive effective interactions. 
The frequency of the lowest breathing 
mode, which can provide an experimental signature of the cross-over,
is calculated.
We predict the existence of a stable 
molecular Tonk-Girardeau gas in the strongly attractive regime.
We believe that our results, which are derived using a 1d model Hamiltonian
with contact interactions,
hold when more realistic interactions are considered
and when the transverse motion
is treated explicitly.

Acknowledgements: GEA, SG and LPP acknowledge support by the Ministero dell'Istruzione, 
dell' Universit\`a e della Ricerca (MIUR). DB acknowledges support by the NSF (grant 0331529).


\begin{thebibliography}{20}
\bibitem{note} The coefficient and sign of the third term on the right hand side of Eq.~(\ref{limit3})
differ from the ones calculated in Ref.~\cite{1DBCS}. This term is crucial to calculate the first correction to $\omega=2\omega_z$ 
for the frequency of the breathing mode in the strong-coupling regime.
\end{thebibliography}
\end{document}